# Evaluation of Crowdsourced Data on Unplowed Roads

Noah J. Goodall[1]

**Abstract**

Transportation agencies routinely collect weather data to support maintenance activities. With the proliferation of smartphones, many agencies have begun using crowdsourced data in operations. This study evaluates a novel unplowed roads dataset from the largest crowdsourced transportation data provider Waze. User-reported unplowed roads in Virginia were compared to national and state weather data for accuracy, and found 81% of reports were near known snow events, with false positives occurring at a regular rate of approximately 10 per day statewide. Reports were largely located on primary roads, limiting the usefulness for transportation agencies who may be most concerned with poorly monitored secondary roads. An effort to encourage unplowed road reporting in Waze through targeted messages on social media did not increase participation. Low reporting may be due to the feature's novelty, recent mild winters, or COVID-19 school and business closures.

---

[1] Senior Research Scientist, Virginia Transportation Research Council, 530 Edgemont Road, Charlottesville, VA 22903. Email: noah.goodall@vdot.virginia.gov. ORCiD: 0000-0002-3576-9886.

## INTRODUCTION
Transportation agencies have used a variety of data to support their operations, such as speed detectors, closed-circuit television (CCTV), and weather sensors. With the proliferation of smartphones and high-speed cellular data, many agencies have begun to leverage data reported either actively or passively by travelers via their mobile phones. Data collected in this manner is often referred to as crowdsourced data.

One of the main providers of crowdsourced transportation data is Waze. Travelers using the Waze smartphone application for navigation have the option to manually enter road conditions and incidents, which are automatically geotagged and pushed to a cloud-based platform (*1*). Other travelers in the vicinity are alerted to the event, and in many cases given the option to confirm that the event is still ongoing.

In 2014, Waze initiated the Connected Citizens Program to provide transportation agencies access to transportation-related user-submitted reports in exchange for agency data (*2*). The Virginia Department of Transportation (VDOT) joined the program in December 2016, sharing real-time incident and work zone data. As part of the effort, VDOT conducted an independent evaluation of the quality of Waze incident and crash data along an urban freeway segment as verified using CCTV images (*3*).

In December 2019 Waze introduced the new data element "unplowed roads." Users could select this incident type to report roads that have not yet been plowed during a snow event. In the Waze app, this appears to users as "Unplowed Roads" as shown in Figure 1. As it appears to VDOT in the Waze feed, this is coded as type "HAZARD" and subtype "HAZARD_WEATHER_HEAVY_SNOW."

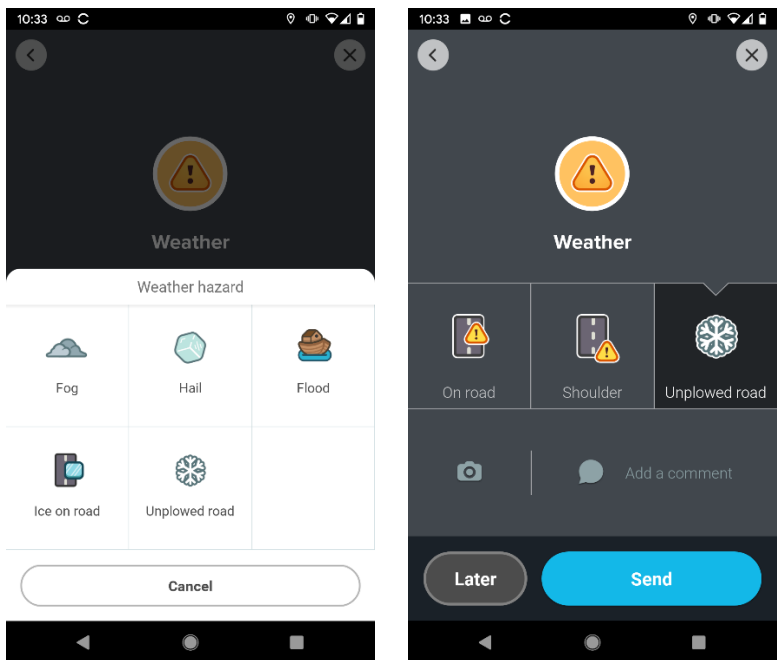

**Figure 1. Screenshot of Waze Application**

These records of unplowed roads may be able to supplement VDOT's Customer Service Center as a method to track the status of plowed roads during a snow event, specifically those roads which are not regularly monitored such as secondary roads. As the reports are generated anonymously by members of the general public using a smartphone application, there may be problems with data quality. Reports are generally cleared after 30 minutes unless other Waze users confirm or deny the initial report, and VDOT does not have access to data on whether a report was confirmed nor how many times it was confirmed. While Waze allows users the option of posting a photo of the report, VDOT does not have access to photos via the Waze feed. Due to these and other issues, the accuracy and coverage of Waze reports should be investigated before integrating Waze unplowed road reports into VDOT's maintenance operations.

The purpose of this study was to evaluate the quality and utility of Waze's unplowed roads data for use in VDOT's snow removal operations. The remainder of the paper is divided into four sections. First, the relevant literature is reviewed. The spatial and temporal coverage of Waze-reported snow events is analyzed and compared to national and state weather data over the 2019/2020 winter season. Waze reports are further analyzed over the 2020/2021 winter season to determine the effect of targeted promotion of the Waze application and comparison with ground truth measurements of plowed and unplowed roads. The final section discusses findings and potential directions for future research.

**LITERATURE REVIEW**
Several studies have investigated the potential for crowdsourced data to supplement transportation agency operations. Silva et al. investigated the characteristics of Waze data and found that reporting volumes were highly skewed, with 35% of users contributing just one report and 0.006% averaging over 100 reports each (*4*). Other studies have used Waze data to predict congestion (*5*), estimate vehicle speeds (*6*), estimate crash rates (*7–10*), and identify secondary crashes (*11*). Evaluations of crowdsourced data quality using Twitter (*12–15*) and Waze (*3*, *16*) have focused largely on disabled vehicles and crashes.

Other studies have attempted to use crowdsourced data in roadway maintenance applications. Praharaj et al. used Waze data to identify flooded roads in a resiliency study (*17*). Several studies have developed smartphone applications that can identify potholes from the phone's accelerometer data (*18*, *19*).

To date, no studies have assessed the accuracy of Waze maintenance data. This study represents the first effort to not only evaluate the accuracy of Waze's unplowed roads data, but also to assess the potential for integrating Waze data into a transportation agency's maintenance operations.

**Planned and Ongoing Unplowed Road Data Uses in Other Cities, States, and Countries**
Over 1,000 transportation agencies have entered into data sharing agreements with Waze, providing public sector traffic and incident data in exchange for real-time access to reports from Waze users. This program is known as Waze for Cities (formerly the Connected Citizens Program). As a member of this group, VDOT surveyed other agencies using the official Waze for Cities email list through Google Groups on April 30, 2020, asking whether any agencies had evaluated or integrated unplowed road data in their operations.

As of May 7, 2020 (seven days later), no agencies had responded, suggesting that agencies may not have yet had a chance to investigate the novel data set. Alternatively, agencies may not have been able to respond due to prioritization of COVID-19-related lockdowns, which were ongoing in most jurisdictions worldwide when the survey was conducted.

**WAZE UNPLOWED ROAD COVERAGE ANALYSIS**

**Methods**
All Waze events with the subtype HAZARD_WEATHER_HEAVY_SNOW over the period of December 10, 2019 (when the data type was added to the Waze app) to March 2, 2020. The Iteris XML feed does not allow queries by incident, but rather by time, date, and road. To access a single record type, all records over a short time period must be queried over a short time period and filtered by incident type offline. To prevent database timeout, queries can request only two-hours' worth of data at a time. To speed the process, VTRC requested only HAZARD_WEATHER_HEAVY_SNOW for Virginia over the selected date ranges from RITIS, who were able to query by event type. The RITIS file included 1,386 reports.

The geographic boundaries of the Waze data from RITIS appear to have been selected by drawing a rectangle around Virginia, and therefore included many reports form Kentucky, Ohio, Delaware, Washington DC, West Virginia, and other states. To filter out non-Virginia reports, the latitudes and longitudes of each report converted to street addresses using a process known as reverse-geocoding. The website Geocodio (*20*) generated the street addresses, including states, from the Waze report latitudes and longitudes. A total of 533 Waze reports were found to be outside Virginia and were removed, leaving 853 Virginia reports.

Reports of unplowed roads were compared with snowfall and snow depth data to evaluate the high-level quality of each report. Reports that occurred on days with no nearby snowfall or snow depth measurements could be classified as false. Snowfall and snow depth data were obtained from the National Oceanic and Atmospheric Administration (NOAA) Daily U.S. Snowfall and Snow Depth records (*21*). Depths were reported at 235 stations, snowfall at 342 stations, with 346 total unique stations reporting. Some stations reported trace values, which were set to 0.01 inches, and many reported missing values for certain days.

The simplest way to determine if a Waze report occurred during a snow event is to check the nearest weather station for reported snowfall on the same day. This approach has several shortcomings. First, snow on the ground may have fallen the previous day, and so the station may report no snowfall for that day. Second, the station may have missing data for that day. Third, a nearby station might not have experienced snowfall even if a nearby report was accurate, for example, because the report occurred at a high elevation and the station was at a low elevation. To obtain a good estimate of the accuracy of Waze snow reports, several nearby stations should be queried for both snowfall and snow depth data.

The following procedure was used to determine whether snow was on the ground at the time and location of a report. For each report, the twenty closest stations were selected in order of ascending great circle distance from the Waze report using the latitudes and longitudes of the stations and the Waze report. The closest station with a snowfall or snow depth measurement greater than 0.01 inches was selected the relevant station. If no stations reported greater than trace amounts (0.01 inches), then the closest station to report trace amounts was selected. If no stations reported trace amounts (0.01 inches) or more, then a measurement of zero inches was used. On average, the third closest station was selected, at a distance of 7 miles from the Waze report. The farthest distance between a Waze report and its selected weather station was 29 miles.

Many stations reported missing data, particularly those in the Northwest region along the I-81 and US-29 corridors during the January 7, 2020 snowstorm. To supplement missing data from NOAA, VDOT's snow event data was downloaded from the Regional Integrated

Transportation Information System's (RITIS) data archive (*22*). In total, 2,838 snow events were logged over the analysis period. Each record includes the event's latitude, longitude, start time, and duration. Events are assigned point locations rather than a geographic areas.

Each Waze record was compared to all VDOT records. Any Waze record that occurred within 20 miles of a VDOT snow report and within a time window of six hours prior to the start of the VDOT report and 24 hours after the end was considered to be valid snow report.

A Waze report that met either the NOAA weather station requirements or the VDOT snow report requirements was considered to have occurred in the presence of snow, and therefore accurate for the purposes of this study.

**Results**

Of the 853 Waze unplowed roads reports received in Virginia between December 10, 2019 and March 2, 2020, proximity to NOAA-recorded snow fall confirmed 343 reports, while proximity to VDOT-reported snow events confirmed 633 reports. The number of Waze reports confirmed either by NOAA or VDOT data was 690 out of 853 (81%).

*Results by Time and Date*

As expected, accurate reports were clustered on days with recorded snow. False reports were evenly distributed on both snow and non-snow days. All days had between 0 and 10 false reports, with an average of 1.9 false reports per day. False reports do not appear to be correlated with true reports, but rather occur at regular frequencies regardless of snow events. Total reports and true reports increase significantly on days with snow events. The only significant snowstorm in Virginia during the winter 2019/2020 analysis period occurred on January 7, which yielded 523 reports, of which 519 (99%) were probably in the presence of snow.

**Table 1. Number of True and False Unplowed Road Reports Per Day**

| Date | Snow | No Snow | Date | Snow | No Snow |
|---|---|---|---|---|---|
| Tuesday, December 10, 2019 | 0 | 1 | Tuesday, January 21, 2020 | 0 | 1 |
| Wednesday, December 11, 2019 | 11 | 2 | Wednesday, January 22, 2020 | 0 | 1 |
| Thursday, December 12, 2019 | 1 | 1 | Thursday, January 23, 2020 | 0 | 1 |
| Friday, December 13, 2019 | 2 | 4 | Friday, January 24, 2020 | 0 | 2 |
| Saturday, December 14, 2019 | 0 | 3 | Saturday, January 25, 2020 | 0 | 2 |
| Sunday, December 15, 2019 | 0 | 3 | Sunday, January 26, 2020 | 0 | 1 |
| Monday, December 16, 2019 | 28 | 0 | Monday, January 27, 2020 | 0 | 3 |
| Tuesday, December 17, 2019 | 0 | 0 | Tuesday, January 28, 2020 | 0 | 0 |
| Wednesday, December 18, 2019 | 0 | 0 | Wednesday, January 29, 2020 | 0 | 2 |
| Thursday, December 19, 2019 | 0 | 3 | Thursday, January 30, 2020 | 0 | 1 |
| Friday, December 20, 2019 | 0 | 0 | Friday, January 31, 2020 | 0 | 1 |
| Saturday, December 21, 2019 | 0 | 2 | Saturday, February 1, 2020 | 0 | 3 |
| Sunday, December 22, 2019 | 0 | 1 | Sunday, February 2, 2020 | 0 | 0 |
| Monday, December 23, 2019 | 0 | 1 | Monday, February 3, 2020 | 0 | 0 |
| Tuesday, December 24, 2019 | 0 | 3 | Tuesday, February 4, 2020 | 0 | 0 |
| Wednesday, December 25, 2019 | 0 | 1 | Wednesday, February 5, 2020 | 0 | 2 |
| Thursday, December 26, 2019 | 0 | 3 | Thursday, February 6, 2020 | 1 | 3 |
| Friday, December 27, 2019 | 0 | 10 | Friday, February 7, 2020 | 7 | 5 |
| Saturday, December 28, 2019 | 0 | 2 | Saturday, February 8, 2020 | 1 | 4 |
| Sunday, December 29, 2019 | 0 | 5 | Sunday, February 9, 2020 | 0 | 0 |
| Monday, December 30, 2019 | 0 | 8 | Monday, February 10, 2020 | 0 | 0 |
| Tuesday, December 31, 2019 | 0 | 1 | Tuesday, February 11, 2020 | 0 | 2 |
| Wednesday, January 1, 2020 | 0 | 1 | Wednesday, February 12, 2020 | 0 | 3 |
| Thursday, January 2, 2020 | 0 | 2 | Thursday, February 13, 2020 | 0 | 5 |
| Friday, January 3, 2020 | 0 | 0 | Friday, February 14, 2020 | 0 | 1 |
| Saturday, January 4, 2020 | 0 | 2 | Saturday, February 15, 2020 | 0 | 1 |
| Sunday, January 5, 2020 | 0 | 1 | Sunday, February 16, 2020 | 0 | 2 |
| Monday, January 6, 2020 | 0 | 2 | Monday, February 17, 2020 | 0 | 5 |
| Tuesday, January 7, 2020 | 519 | 4 | Tuesday, February 18, 2020 | 0 | 1 |
| Wednesday, January 8, 2020 | 21 | 1 | Wednesday, February 19, 2020 | 0 | 1 |
| Thursday, January 9, 2020 | 3 | 0 | Thursday, February 20, 2020 | 22 | 4 |
| Friday, January 10, 2020 | 1 | 1 | Friday, February 21, 2020 | 4 | 1 |
| Saturday, January 11, 2020 | 0 | 3 | Saturday, February 22, 2020 | 1 | 0 |
| Sunday, January 12, 2020 | 0 | 0 | Sunday, February 23, 2020 | 0 | 1 |
| Monday, January 13, 2020 | 0 | 1 | Monday, February 24, 2020 | 0 | 0 |
| Tuesday, January 14, 2020 | 0 | 0 | Tuesday, February 25, 2020 | 0 | 2 |
| Wednesday, January 15, 2020 | 0 | 1 | Wednesday, February 26, 2020 | 0 | 4 |
| Thursday, January 16, 2020 | 1 | 0 | Thursday, February 27, 2020 | 1 | 4 |
| Friday, January 17, 2020 | 2 | 0 | Friday, February 28, 2020 | 1 | 4 |
| Saturday, January 18, 2020 | 57 | 1 | Saturday, February 29, 2020 | 3 | 10 |
| Sunday, January 19, 2020 | 1 | 1 | Sunday, March 1, 2020 | 0 | 1 |
| Monday, January 20, 2020 | 2 | 3 | Monday, March 2, 2020 | 0 | 1 |

*Results by Road Type*
Waze reports can be created by users at any time and are not limited to public roads. Other Waze reports may cover Interstate highways, where VDOT has generally good surveillance and plowing coverage. Most of the value from Waze unplowed road reports are expected on minor streets, rural routes, and neighborhoods that plows might miss or be unable to service during a storm.

Table 2 shows the numbers of true (snow likely present) and false (no snow likely present) reports by Waze-reported road type. Waze uses a proprietary method to classify road type. There is no correlation to VDOT-maintained roads, so that the category "streets," for example, covers surface streets in both VDOT-maintained and municipality-maintained jurisdictions. In addition, 17% of roads had no classification. Finally, a spot check noted some errors, for example "ramps" where the geospatial coordinates were not ramps. Caution is warranted when interpreting these results.

The road types of most interest to VDOT are highlighted in bold in Table 2. These roads (streets, primary streets, secondary, and ramps) accounted for 252 true reports, 37% of the total true reports. Of the 252 true reports recorded over the study period, 197 (78%) were from January 7, 2020.

**Table 2. Waze Unplowed Road Reports by Road Type and Actual Presence of Snow**

| Road Type | Snow Present | | No Snow Present | |
|---|---|---|---|---|
| Primary | 180 | *(26%)* | 29 | *(18%)* |
| Freeways | 128 | *(19%)* | 64 | *(39%)* |
| Streets | **89** | *(13%)* | 17 | *(10%)* |
| Primary Streets | **83** | *(12%)* | 12 | *(7%)* |
| Secondary | **70** | *(10%)* | 12 | *(7%)* |
| Ramps | **10** | *(1%)* | 6 | *(4%)* |
| Parking lot road | 7 | *(1%)* | 3 | *(2%)* |
| Private road | 4 | *(<1%)* | 1 | *(<1%)* |
| Blank | 119 | *(17%)* | 19 | *(12%)* |
| Total | 690 | *(100%)* | 163 | *(100%)* |

*Results by Hour and Road Type*
Only the storm on January 7, 2020 generated enough Waze reports to analyze the number of new reports generated per hour. Figure 1 shows the number of new Waze unplowed road reports generated each hour during and after the storm, categorized by road type. The road types of most interest to VDOT (streets, primary streets, secondaries, and ramps) produced between 10 to 20 new reports each per hour. Combined, these accounted for 10 to 50 reports per hour during the peak of the storm. Reports reduced sharply following the storm, with fewer than five new reports per road type by 5 PM (17:00). The data from the January 7 storm suggests that Waze users may not be reporting unplowed roads in large numbers in the hours and days following the storm, although it should be noted that the January 7 storm was minor with fewer than six inches of total accumulation in Virginia (*23*).

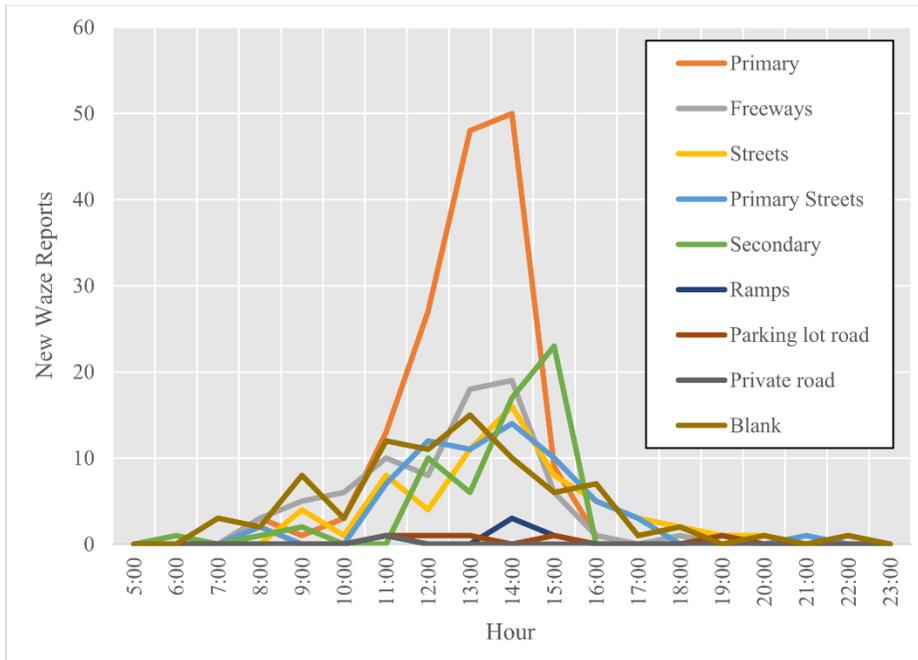

**Figure 2. New Unplowed Road Reports in Waze per Hour by Road Type, January 7, 2020**

*Spatial Analysis*

Similar to most other categories of Waze reports, crowdsourced data is heavily concentrated along major corridors and in urban regions with high-volume peak period travel. Figures 2 and 3 show true ("Snow") and false ("No Snow") reports from the January 7 storm and all other dates. Reports are concentrated along major Interstate corridors such as I-81, I-66, I-64, and I-95, as well as some high-volume primary corridors such as US-29.

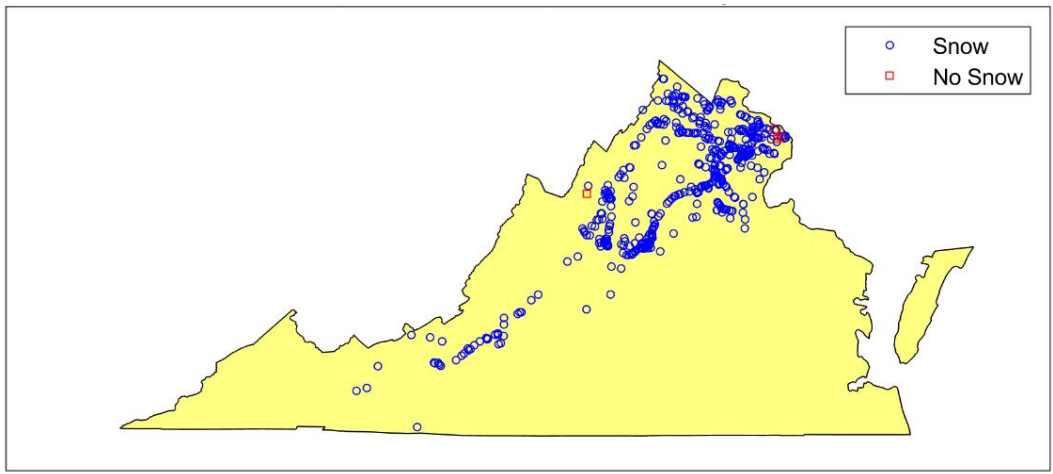

**Figure 3. Waze Unplowed Road Reports, January 7, 2020**

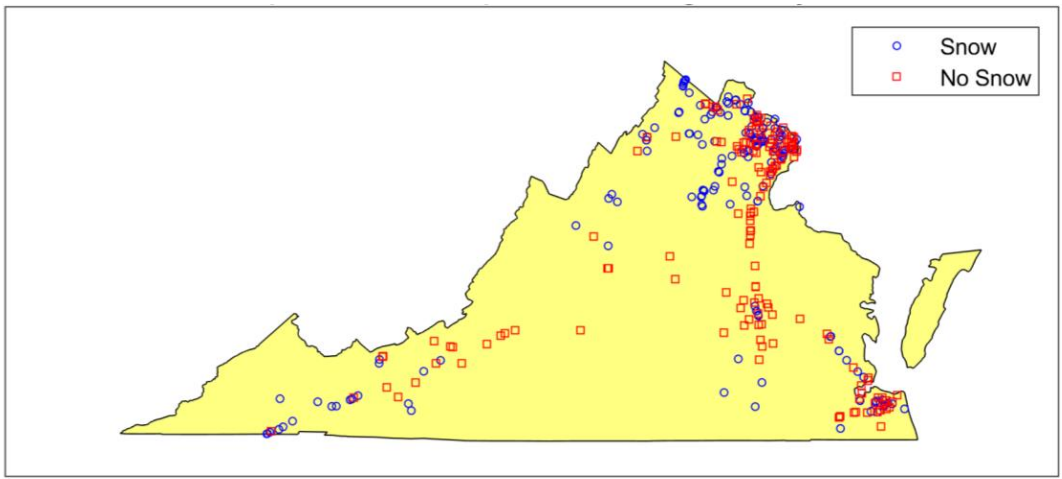

**Figure 4. Waze Unplowed Road Reports, December 10, 2019 – March 2, 2020 Excluding January 7, 2020**

A closer analysis reveals that even in urban areas, most of the reports are from high-volume roadways. Figure 4 displays a sample of Waze reports from January 7, 2020 in parts of Arlington and Fairfax County. Most of the reports occur on major roads such as US-50, US-28, US-29, SR 267, and I-66. Only a few occur in neighborhoods or on surface streets.

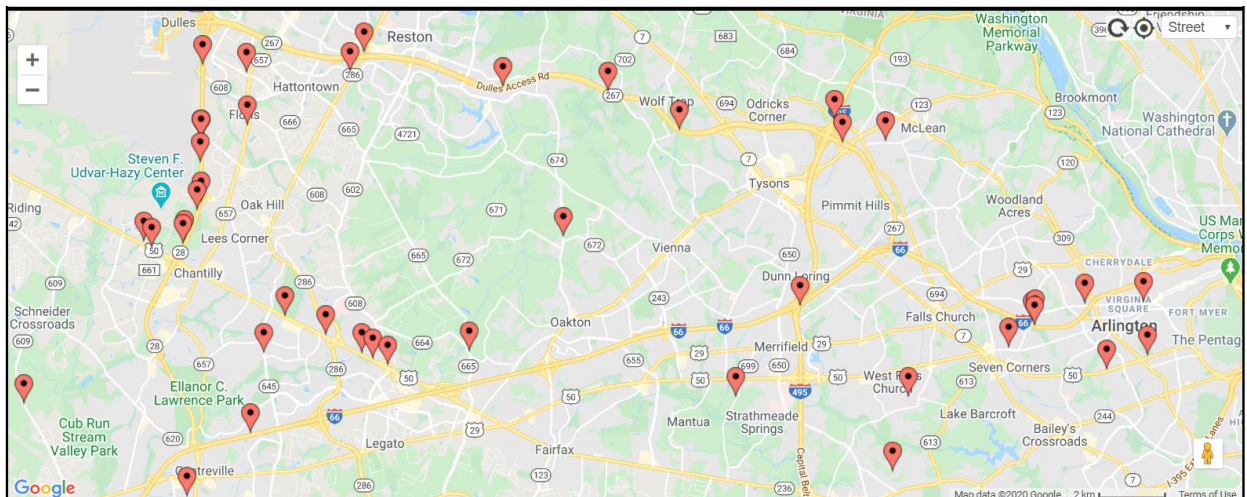

**Figure 5. Sample of Waze Unplowed Road Reports in Arlington and Fairfax Counties, January 7, 2020**

**EFFECT OF PROMOTION ON WAZE PARTICIPATION**

If more residents use the Waze app to report unplowed roads in their neighborhood, transportation agencies may be able to use this data to assist snow plow operations and target unserved or underserved streets and neighborhoods. In order to encourage Waze reporting by residents in their homes, VDOT sent messages to users of the social networking site Nextdoor (*24*) residing in the Bristow Village neighborhood at 8:15 AM on January 31, 2021. The message notified residents that VDOT would be monitoring Waze unplowed road reports and provided a link to a tutorial on the Waze feature. Messages were sent near the beginning of the snow event. Figure 6 shows screenshots of the messages.

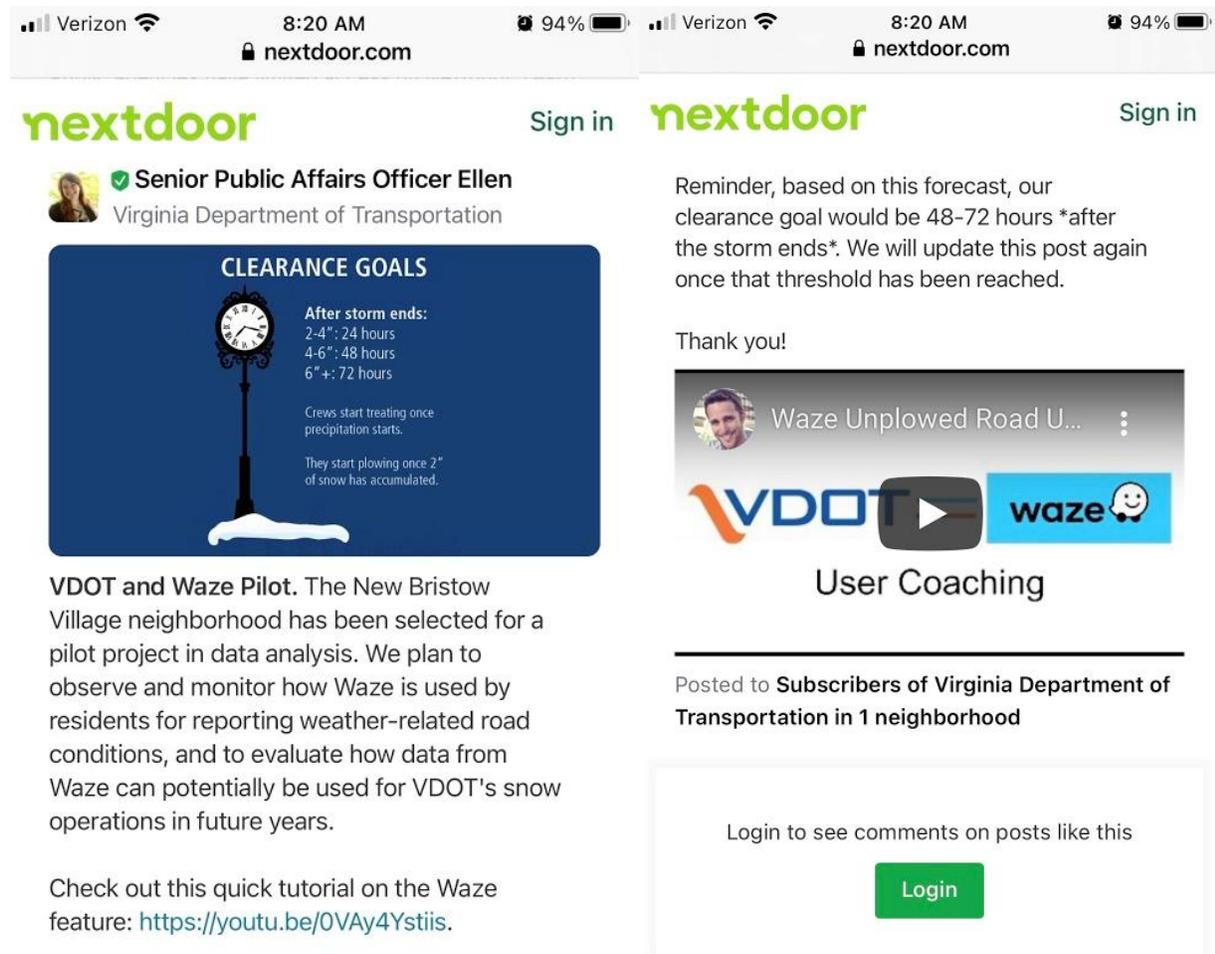

**Figure 6. Screenshots of messages sent on social networking platform Nextdoor to 395 residents of the Bristow Village neighborhood at 8:15 AM on January 31, 2021.**

    The original intent of the study was to compare the number of reports originating from the Bristow Village neighborhood to similar neighborhoods in Chantilly, Virginia that experienced similar weather but were not messaged on Nextdoor. In fact, only 38 Waze unplowed road reports were received in the 54 hours following the first snowfall in all of Northern Virginia. Because of low reporting volumes, other neighborhoods could not be used as a baseline.

**Results**
Waze unplowed road reports did not appear to increase as a result of the promotional messages. No Waze alerts were reported at all in the targeted neighborhood, nor were any snow or icy roads reported on the Waze app in the entire Northern Virginia region prior to the neighborhood's roads reported as cleared by plow staff at 2:00 PM.

The lack of reports could be due to a fairly light snowstorm with low accumulation and quick response by snow removal operations. The snowstorm also occurred on a Sunday when fewer drivers needed to get to work or school, and so residents may have had less urgency in reporting a road as unplowed. Finally, data collection occurred during widespread school and business closures associated with the COVID-19 distancing restrictions, which may have further reduced the urgency to report unplowed roads.

**CONCLUSIONS AND DISCUSSION**
The purpose of this study was to evaluate the volume, scope, and quality of Waze's crowdsourced unplowed roads data, and to determine whether targeted reminders of the Waze reporting too could increase use among residents.

The quality of Waze data appears sufficient to warrant further study. Between 3 and 10 reports per day came from times and locations where there was no evidence of snow. This figure stayed consistent during snowstorms, when over 500 reports were received yet still only 3 to 10 were in areas without snow. It could not be determined from this effort whether the reports from areas with snow were on roads that had already been cleared.

The amount of Waze data generated appeared limited by a light snow season and COVID-19 travel restrictions. Only one significant snowstorm occurred during the 2019/2020 winter, on January 7, 2020. This was only 28 days after the feature was introduced in Waze, and users may have been unaware of it. Also, it was a minor snowstorm with accumulation between 3 and 6 inches in northwest and northern Virginia. In total 519 reports were collected from snow areas during the storm, but only 197 were from roads that were of value to Virginia, e.g. streets, primary streets, secondary, and ramps. Finally, Waze reports dropped off after the initial storm, suggesting that users did not continue to report unplowed roads in the hours following the initial snowfall. These rates could increase significantly with more awareness of the feature, active promotion of the feature, and heavier snowfalls. The 2020/2021 winter season also experienced only minor snow accumulation, and widespread school and business closures associated with the ongoing COVID-19 outbreak caused reduced mileage driven resulting in fewer Waze reports.

Future research efforts may reevaluate Waze report volume and coverage during a more severe winter event, as recent winters in Virginia have been mild. As COVID-19 restrictions begin to ease, traffic volumes will likely increase along with Waze reporting. This may also allow researchers to obtain more conclusive results.

**ACKNOWLEDGEMENTS**
This work was sponsored by the Virginia Department of Transportation and the Federal Highway Administration (FHWA) State Transportation Innovation Council (STIC) Incentive program. Amy McElwain, Candice Gibson, Larry Camp, Nathan O'Kane, and Hari Sripathi supported the study design and performed on-the-ground data collection. Jenni McCord and Ellen Kamilakis conducted public outreach.

**AUTHOR CONTRIBUTION STATEMENT**
N.J. Goodall was sole author and is responsible for the entirety of this paper.